\documentclass[preprint,showpacs,preprintnumbers,amsmath,amssymb,prb]{revtex4}

\usepackage{amsmath,amssymb}
\usepackage{epsfig}

\def\e{\begin{equation}}
\def\f{\end{equation}}
\def\=#1{\overline{\overline{#1}}}
\def\_#1{{\bf #1}}
\def\o{\omega}
\def\E{\varepsilon}
\def\M{\mu}
\def\.{\cdot}

\def\l#1{\label{eq:#1}}
\def\r#1{(\ref{eq:#1})}
\def\d{\partial}

\def\l#1{\label{eq:#1}}
\def\r#1{(\ref{eq:#1})}

\begin{document}

%\preprint{APS/123-QED}

\title{Non-local permittivity from a quasi-static model for a class of wire media}

\author{Stanislav I. Maslovski}
 \email{stas@co.it.pt}
\author{M\'{a}rio G. Silveirinha}
\affiliation{
Departamento de Engenharia Electrot\'{e}cnica\\
Instituto de Telecomunica\c{c}\~{o}es, Universidade de Coimbra\\
P\'{o}lo II, 3030-290 Coimbra, Portugal}

\date{\today}

\begin{abstract}
  A simple quasi-static model applicable to a wide class of wire media
  is developed that explains strong non-locality in the dielectric
  response of wire media in clear physical terms of effective
  inductance and capacitance per unit length of a wire. The model is
  checked against known solutions and found to be in excellent agreement
  with the results obtained by much more sophisticated analytical and
  numerical methods. Special attention is given to suppression of the
  spatial dispersion effects in wire media.
\end{abstract}


%\pacs{Valid PACS appear here}% PACS, the Physics and Astronomy
                             % Classification Scheme.
%\keywords{Suggested keywords}%Use showkeys class option if keyword
                              %display desired
\maketitle

\section{Introduction}

Wire media are structured materials formed by many conducting wires
embedded in a host medium. The wires are normally considered to be
very long compared to the wavelength in the host medium, but the
diameter of the wires is only a small fraction of the lattice
constant. The known analytical models of wire media
\cite{Pendry_plasmons_PRL_1996,Pendry_plasmons_JPCM_1998,
Belov_wiremedium_JEWA_2002,Maslovski_quasistatic_MOTL_2002,Efros_WM_PRB_2002,
Shvets_wires_PSPIE_2003,Belov_dispersion_PRB_2003,Constantin_WM,
Silveirinha_3dconnected_I3EMTT_2005,Silveirinha_ENG_Plasmonic_2006,
Silveirinha_crosswires_PRB_2009} treat them as crystals of
infinitely long conducting cylinders. The cylinders may be arranged
in different types of lattices resulting in different types of
anisotropy of the wire crystals. It is known that wire media may
exhibit strong spatial dispersion, so that the permittivity dyadic
$\=\E(\omega, \_k)$ in such media depends on both frequency and wave
vector. For instance, the permittivity dyadic of uniaxial wire
medium with one set of thin ideally conducting wires oriented along
$\_z_0$ reads \cite{Belov_dispersion_PRB_2003} \e{\=\E(\o, \_k)\over
  \E_0} = \=I_{\rm t} + \left(1 - {k_{\rm p}^2 \over k_0^2 -
    k_z^2}\right)\_z_0\_z_0, \l{epsilon} \f where
$k_0=\omega\sqrt{\E_0\M_0}$, $\E_0$ and $\M_0$ are the permittivity
and the permeability of the host medium, $k_{\rm p}$ is the plasma
wavenumber, $k_z$ is the wave vector component along $\_z_0$, and
$\=I_{\rm t}$ is the unit dyadic in the plane orthogonal to~$\_z_0$.

It is well known that the wire medium supports propagation of
transverse electromagnetic modes (TEM) which are basically the modes
of a multi-wire transmission line.\cite{Belov_dispersion_PRB_2003,
Silveirinha_ENG_Plasmonic_2006} Such modes propagate along the
wires with the velocity equal to the speed of light in the host
medium. The distribution of the microscopic $\_E$ and $\_H$ fields
associated with the TEM modes is static-like in the planes orthogonal
to the wires, with the electric force lines
emerging from and ending at the surfaces of the wires. It can be
easily proven that there is electrical charge accumulated on the
wires associated with these modes. In Ref.~\onlinecite{Maslovski_disser_2004} it was shown (for the
uniaxial wire medium case) that when this charge and the related
potential are taken into account it is possible to obtain Eq.~\r{epsilon}
from simple quasi-static considerations similar to
those used in Ref.~\onlinecite{Maslovski_quasistatic_MOTL_2002}. Thus, it was
shown that the strong spatial dispersion in wire media can be
correctly described in a quasi-static approximation. In this paper
we extend these considerations to a wide class of wire media, and
propose an analytical model based on the effective inductance and
capacitance per unit length of a wire.

The other motivation for this study is the suppression of the
nonlocal effects in wire media. In a recent paper by
Demetriadou {\it et al.}\cite{Demetriadou_taming_JPCM_2008} the charge accumulated on the
wires together with the rather small capacitance of thin wires were
identified as the reasons for the spatial dispersion in wire mesh: a
metamaterial formed by three sets of wires oriented along three
Cartesian coordinate axes and joined at the crossing points. A
rigorous analytical model of such medium was developed
in Refs.~\onlinecite{Silveirinha_3dconnected_I3EMTT_2005,Silveirinha_crosswires_PRB_2009}.
The authors of Ref.~\onlinecite{Demetriadou_taming_JPCM_2008} make use of this
model and full wave simulations to justify their main claims. They
also propose certain ways how to decrease the spatial dispersion
effects. The basic idea is to increase the capacitance of the wires
by periodically loading them with metallic bodies or patches, or
alternatively to increase the inductance per unit length by coating
the wires with a magnetic material. Somehow related to this work, it
was shown in Refs.~\onlinecite{Alex_Mushrooms_MTT_2009,
Olli_Mushrooms_MTT_2009} that for a substrate formed by a wire
medium slab capped with an array of patches (the so-called mushroom
substrate \cite{Sievenpiper_Mushrooms_MTT_1999}) the response of the
wire medium is essentially local. A different strategy to reduce the
spatial dispersion was reported in Ref.~\onlinecite{Silveirinha_crosswires_PRB_2009}, where it was shown that at
infrared frequencies the plasmonic properties of metals may enable
the design of artificial plasmas that mimic more closely a
continuous local isotropic medium with negative permittivity.

In this work, we generalize the theories reported in previous
studies,
\cite{Pendry_plasmons_PRL_1996,Pendry_plasmons_JPCM_1998,Belov_wiremedium_JEWA_2002,Maslovski_quasistatic_MOTL_2002,Shvets_wires_PSPIE_2003,Belov_dispersion_PRB_2003,
Efros_WM_PRB_2002, Silveirinha_ENG_Plasmonic_2006,
Silveirinha_3dconnected_I3EMTT_2005, Constantin_WM,
Silveirinha_crosswires_PRB_2009} and propose a quasi-static
homogenization model that accurately characterizes the nonlocal
dielectric function of a wide class of wire media (both arrays of
parallel wires, and arrays of connected wires), including the case
where the wires are periodically loaded with conducting metallic
bodies. In particular, we demonstrate that our analytical theory
models accurately the electric response of a uniaxial wire medium
loaded with patches, and we discuss the physics of the suppression
of spatial dispersion in such structures.

\section{Uniaxial wire medium \label{SecWires}}

\label{uniaxial} We will start with the simplest possible case of
the uniaxial wire medium with one set of wires oriented along the
$z$-axis. We will follow the treatment presented in
Ref.~\onlinecite{Maslovski_disser_2004}.

We are interested in the longitudinal (\emph{zz}) component of the
permittivity dyadic. To get an expression for it in the quasi-static
limit we assume that the radius of the wires $r_0$ and the distance
between the wires (the lattice period) $a$ are much less than the
wavelength in the medium. Let us note that for the model we are
going to develop the exact arrangement of the wires is not
important, it is just enough to know the average distance between a
pair of neighboring wires in a structure.

\begin{figure}[htb]
  \centering \epsfig{file=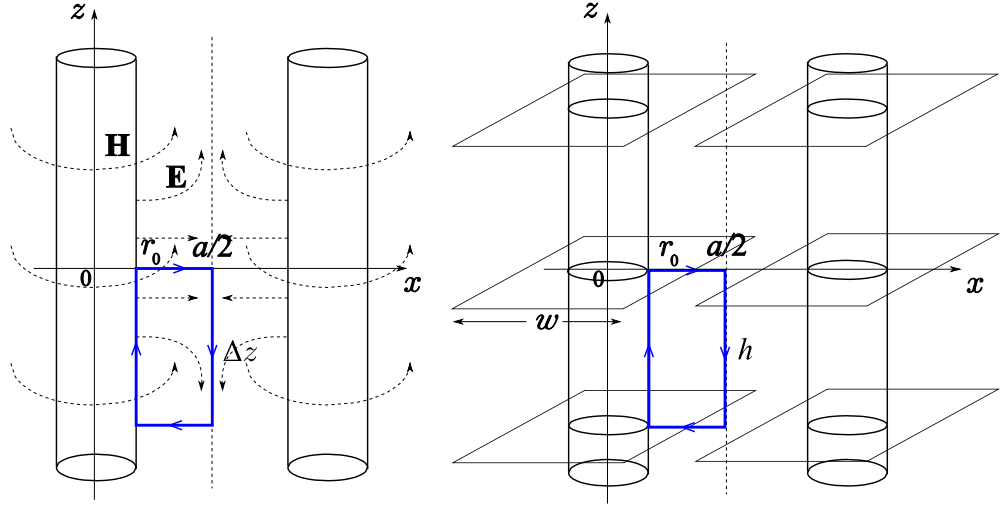,width=0.8\textwidth}
  \caption{\label{fig1} (Color online) A pair of wires of the uniaxial wire medium
    without (on the left) and with patches (on the right). The
    integration path used to define Eq.~\r{circulation} is shown by
    the blue rectangular contour.}
\end{figure}

Denoting the average (macroscopic) electric field along $z$ axis in
the medium by $\langle E_z\rangle$, one can write the following
relation between this field component and the current in the
wires~$I_z$: \e \langle E_z \rangle = (j\o L + Z_w) I_z + {\d\varphi\over
\d z}, \l{main_uni} \f where $L$ is the effective inductance per
unit length of the wire, $Z_w$ is the self-impedance of the wire per
unit length which accounts for the finite conductivity of metallic
wires at microwave frequencies or plasmonic behavior at optical
frequencies, and $\varphi$ is the additional potential due to
charges on the wires.

This relation can be obtained integrating the microscopic electric
field over a path shown in Fig.~\ref{fig1}. The path goes first
along the surface of a wire then to the middle line in a pair of two
neighboring wires, then along this middle line and, finally, back to
the surface of the wire. The circulation of the microscopic electric
field $\_E(x,z)$ over this path reads
\begin{multline}
\oint\_E\.\_{dl} = \int\limits_{z}^{z+\Delta z}\!\!E_z(r_0, z')\,dz'
-\!\!\int\limits_{z}^{z+\Delta z}\!\!E_z(a/2, z')\,dz'
+\int\limits_{r_0}^{a/2}E_x(x,z+\Delta z)\,dx
-\int\limits_{r_0}^{a/2}E_x(x,z)\,dx. \l{circint}
\end{multline}
The first integral in this relation represents the voltage drop
along the surface of the wire and, therefore, can be expressed in
terms of the wire current and the wire self-impedance per unit
lenght.  The second integral is the voltage drop along the symmetry
line shown in Fig.~\ref{fig1}. In the same manner as it was done in
Ref.~\onlinecite{Maslovski_quasistatic_MOTL_2002} we relate this voltage drop
with the macroscopic electric field in the medium. After doing this
the circulation of the electric field reads (when $\Delta z$ is
small enough) \e \l{circulation} \oint\_E\.\_{dl} = (Z_wI_z -
\langle E_z\rangle)\Delta z +\varphi(z + \Delta z) - \varphi(z),
\quad\mbox{where } \varphi(z) = \int\limits_{r_0}^{a/2}\!E_x(x,z)\,
dx. \f The electric field circulation equals minus the time
derivative of the magnetic flux that penetrates the area bounded by
the integration path: $\oint \_E\.\_{dl} = -j\o\Phi = -j\o L I_z
\Delta z$, from which we immediately get \r{main_uni} when $\Delta
z\rightarrow 0$.

In general, the effective inductance $L$ depends on the specific
microstructure of the system (e.g. if the wires are coated or not
with some material). In the particular case in which the wires are
conducting cylinders (with no material coating), it was
shown,\cite{Maslovski_quasistatic_MOTL_2002} by calculating the magnetic
flux of a pair of neighboring wires in the quasi-static
approximation, that $L$ verifies: \e L = {\M_0\over
2\pi}\log{a^2\over
  4r_0(a-r_0)}. \l{inductance}\f
It may be verified that the above formula also applies to the case
where the wires are loaded with metallic patches (Fig. \ref{fig1}, right).

The additional potential caused by the charges on the wires can be
found by placing a linear charge density $\rho$ on the wires and by
calculating the corresponding electrostatic potential $\varphi$
created by the fluctuating part of the microscopic electric field.
Thus $\rho$ is responsible for the electric field component
orthogonal to the wires. We introduce an effective capacitance $C$
per unit length, such that it verifies: \e \varphi(z) =
{\rho(z)\over C}. \l{defcapacitance} \f Notice that the considered
capacitance is calculated by placing an $identical$ linear charge
density over the wires (differently from the traditional definition
of capacitance, which assumes that charge density over two
conductors is antisymmetric). In the same manner as the inductance, the
capacitance depends on the microstructure of the system. In the
quasi-static limit a pair of charged wires (with no attached
conducting bodies) induces the field (see Fig.~\ref{fig1}) \e E_x =
{\rho\over
  2\pi\E_0}\left[{1\over x} - {1\over a-x}\right].  \f This expression
has the same form as the one used in Ref.~\onlinecite{Maslovski_quasistatic_MOTL_2002} for the quasi-static magnetic
field of a pair of lines of current. Therefore, for this particular
case the capacitance is given by \e  {1\over C} = {1\over
  2\pi\E_0}\log{a^2\over 4r_0(a-r_0)}. \l{capacitance} \f The
capacitance for a system of wires loaded with conducting patches
(Fig.~\ref{fig1}, right) is calculated in Appendix A.

Considering now a monochromatic plane wave of current excited in the
crystal, the currents in the wires can be written in the form \e
I_z(z) = I_0 e^{-j k_z z}, \f and thus the linear density of the
charge associated with the currents verifies \e \rho(z) = -{1\over
j\o} {dI_z(z)\over dz} = {k_z\over\o}I_z(z). \f These charges are
responsible for the electric field component orthogonal to the
wires.

Hence, the relation \r{main_uni} can be rewritten in terms of the
effective inductance and of the effective capacitance per unit
length of the wire as \e \langle E_z\rangle = \left(j\o
L + Z_w + {k_z^2\over j\omega C}\right)I_z. \f Already in this
expression one can identify the spatial dispersion term proportional
to the square of the $z$-component of the wave vector.

The macroscopic polarization current in wire media is the average of
the currents in separate wires.  Let $A_{\rm cell}$ be the average area
in the $xy$ plane per one wire of the crystal.  Then the macroscopic
polarization current is $J_z = I_z/A_{\rm cell}$. The macroscopic
displacement field is $D_z = \E_0 \left\langle {E_z } \right\rangle
 + J_z/(j\o)$. Therefore,  after some
algebra we find that the longitudinal component of the permittivity
dyadic is given by
\e {\E_{zz}\over \E_0} = 1 - {k_{\rm p}^2\over
k_0^2 - j\xi k_0 - k_z^2 /
  n^2}, \l{uniaxial_permittivity} \f where $k_{\rm p}^2 =
\M_0/(A_{\rm cell}L)$, $n^2 = LC/(\E_0\M_0)$, $\xi =
(Z_w/L)\sqrt{\E_0\M_0}$. It may be easily checked that the above
formula reduces to Eq.~\r{epsilon} in the case of perfectly
conducting straight wires ($Z_w=0$) [also, for unloaded wires $n = 1$ as
is seen from Eqs.~\r{inductance} and \r{capacitance}]. More
generally, when the wires are characterized by the complex
permittivity $\E_0 \E_{\rm m}$ (e.g., thin plasmonic rods at optical frequencies), the
impedance $Z_w$ is given by, \e Z_w = \frac{1}{j\o \pi
r_0^2\E_0(\E_{\rm m} - 1)}, \l{Zw}\f where $r_0$ is the radius of the
rods. It may be easily verified that in this scenario Eq.
\r{uniaxial_permittivity} reduces to formula (16) of Ref.~\onlinecite{Silveirinha_ENG_Plasmonic_2006},
which was calculated using a local field based approach. Thus, Eq. \r{uniaxial_permittivity}
generalizes the previous homogenization models of the uniaxial wire
medium.

Nevertheless, it is worth noting that the expression for the plasma
wavenumber obtained in the present paper differs from the one
derived in previous works.~\cite{Belov_wiremedium_JEWA_2002,
Silveirinha_ENG_Plasmonic_2006} Namely, under the approach
developed above we have \e (k_{\rm p}a)^2 = {2\pi \over
\log{a^2\over 4r_0(a - r_0)}}. \l{myplasm} \f
In Refs.~\onlinecite{Belov_wiremedium_JEWA_2002, Silveirinha_ENG_Plasmonic_2006},
under a thin wire approximation, it was obtained that \e (k_{\rm
p}a)^2 \approx {2\pi \over 0.5275 + \log{a\over 2\pi r_0}}.
\l{belovplasm} \f

One can notice that \r{belovplasm} gives unphysical results for any
${r_0/a} \ge (2\pi)^{-1}\exp(0.5275) \approx 0.27$. Contrary,
Eq.~\r{myplasm} gives a physically sound result in the limit $r_0
\rightarrow a/2$ when the surfaces of two wires touch: It predicts
an infinite growth in the magnitude of $k_{\rm p}$ in this limit. It
can be also checked numerically that the accuracy of \r{myplasm} is
better than \r{belovplasm} when $r_0 \approx 0.1a$ or larger,
whereas the opposite behavior is observed for $r_0 < 0.05a$.
Nevertheless, both formulas have the same asymptotic behavior when
$r_0\rightarrow 0$. At $r_0/a = 0.05$ (this ratio has been used in
our numerical simulations that are discussed in
Section~\ref{supression}) the formulas \r{myplasm} and
\r{belovplasm} overestimate the plasma frequency by about 3\%.

Another asymptotic expression for the normalized plasma frequency
which is often cited was obtained in
Refs.~\onlinecite{Pendry_plasmons_PRL_1996,Pendry_plasmons_JPCM_1998} but even
for rather small wire radii its accuracy is worse than that of
\r{myplasm} and \r{belovplasm}. Also, it does not predict the
infinite growth of $k_{\rm p}$ when $r_0\rightarrow a/2$.

It should be emphasized that Eq.~\r{uniaxial_permittivity} is in
principle valid for a wide class of wire media (e.g. wires with
attached conducting bodies). The parameters
$C$ and $L$ depend on the specific microstructure of the system. The
magnitude of the spatial dispersion term $k_z^2/n^2$ in
\r{uniaxial_permittivity} can be reduced by increasing the value of
$n = \sqrt{LC/(\E_0\M_0)}$. This quantity has the meaning of slow-wave factor for quasi-TEM
waves propagating along the wires. As mentioned before, for unloaded
straight wires $n = 1$. As discussed in Ref.~\onlinecite{Demetriadou_taming_JPCM_2008}, the capacitance $C$ can be
increased by loading wires with metallic patches and the inductance
$L$ can be increased by placing wires in ferromagnetic shields. An
alternative way to increase the inductance is to use helices instead
of straight wires. Associated bi-anisotropy in helix medium can be
compensated if both right- and left-handed helices are used.

Attaching metallic or dielectric bodies to the wires also changes
the transversal components of the permittivity dyadic. We will study
this effect with more details in Section~\ref{sect_uniax_supr}.

\section{Wire mesh}
\label{mesh} The (3D) wire mesh is a wire crystal formed by three
mutually orthogonal sets of wires joined at the intersection points.
The electromagnetics of such metamaterial have been studied in
several recent works.
\cite{Hudlicka_WM3D_PIER_2006,Shapiro_WM3D_OL_2006,Silveirinha_3dconnected_I3EMTT_2005,Silveirinha_crosswires_PRB_2009}
In the following derivation we assume a cubic lattice, but after a
straightforward generalization the same method can be applied to
structures of more complex geometries. Similar to the case studied
in section \ref{SecWires}, metallic or dielectric bodies may be
attached to the wires.

In the wire mesh we get three components of the polarization current
related with the currents in three orthogonal sets of wires. The
currents in the wires are related to the average electric field in
the medium in a manner similar to the uniaxial case:
\begin{eqnarray}
\l{wiremesh1} \left\langle {E_x } \right\rangle &=& (j\o L + Z_w) I_x + {\d\varphi\over \d x},\\
\left\langle {E_y } \right\rangle &=& (j\o L + Z_w) I_y + {\d\varphi\over \d y},\\
\l{wiremesh3} \left\langle {E_z } \right\rangle  &=& (j\o L + Z_w)
I_z + {\d\varphi\over \d z}.
\end{eqnarray}
Because the wires are joined at the crossing points they are locally
under the same potential, that is why we have the same $\varphi$ in
all three equations. But the currents in three sets of wires can
differ and that is taken into account by the variables $I_x$, $I_y$,
and $I_z$.

Let us consider a unit cell of the wire mesh with three intersecting
connected wires.  The total charge $q$ accumulated on these three
wires per unit cell can be found as \e q = -{a\over
  j\o}\left({dI_x\over dx} + {dI_y\over dy} + {dI_z\over dz}\right).
\f Because the wires are electrically connected and their effective
capacitance per unit length is the same, this charge is equally
distributed among the three wires in the unit cell. Therefore, for the
linear charge densities on the wires we have in a vicinity of the unit
cell \e \rho_x = \rho_y = \rho_z = {q\over 3a} = -{1\over
  3j\o}\left({dI_x\over dx} + {dI_y\over dy} + {dI_z\over dz}\right).
\f Using the same notation for the effective capacitance of a wire as
above we can write the potential $\varphi$ as \e \varphi = -{1\over
  3j\o C}\left({dI_x\over dx} + {dI_y\over dy} + {dI_z\over dz}\right)
= {1\over 3\omega C}\left(k_x I_x + k_y I_y + k_z I_z\right), \f
where we have taken into account that the currents on the wires
change on average as \e I_{n}=I_{n}^0 e^{-jk_{n} n}, \quad n=x,y,z.
\f Now we can substitute this expression for the additional
potential into \r{wiremesh1}--\r{wiremesh3}.  Doing this we obtain
the following system of equations:
\begin{eqnarray}
\left\langle {E_x } \right\rangle &=& (j\o L + Z_w + {k_x^2\over 3j\o C}) I_x + {k_x\over 3j\o C}(k_yI_y + k_zI_z),\\
\left\langle {E_y } \right\rangle &=& (j\o L + Z_w + {k_y^2\over 3j\o C}) I_y + {k_y\over 3j\o C}(k_zI_z + k_xI_x),\\
\left\langle {E_z } \right\rangle &=& (j\o L + Z_w + {k_z^2\over
3j\o C}) I_z + {k_z\over 3j\o C}(k_xI_x + k_yI_y).
\end{eqnarray}
By introducing a vector of currents $\_I =
I_x\_x_0+I_y\_y_0+I_z\_z_0$ we rewrite this system in a more compact
form using dyadics: \e \left\langle { \_E } \right\rangle =
\left[(j\o L +Z_w)\=I + {\_k\_k\over 3j\o C}\right]\.\_I,
\l{electric_field} \f where $\=I$ is the unit dyadic and ${\bf{kk}}
\equiv {\bf{k}} \otimes {\bf{k}}$ is the dyadic (tensor) product of
two vectors. Now it is only a matter of inverting the dyadic in brackets of
\r{electric_field} to get the permittivity dyadic of the wire mesh.

The average polarization in the medium is $\_P = \_I/(j\o A_{\rm cell})
+ \_P_{\rm t}$, where $\_P_{\rm t}$ accounts for additional
polarization due to finite thickness of the wires or metallic bodies
attached to the wires. For a crystal of cubic symmetry we can write
$\_P_{\rm t} = \E_0(\E_{\rm t}-1) \left\langle { \_E }
\right\rangle$, therefore the displacement vector $\_D = \E_0\E_{\rm
t}\langle \_E\rangle + \_I/(j\o A_{\rm cell})$, and \e {\=\E(\o,\_k)\over \E_0} =
\E_{\rm t}\=I + {1\over j\omega\E_0 A_{\rm cell}}\left[(j\o L +Z_w)\=I +
  {\_k\_k\over 3j\o C}\right]^{-1}, \f or, after some dyadic algebra,
\e {\=\E(\o,\_k)\over \E_0} = \left(\E_{\rm t} - {k_{\rm p}^2\over
    k_0^2 - j\xi k_0}\right)\=I - {k_{\rm p}^2\,\_k\_k\over
  3n^2[k_0^2-j\xi k_0][k_0^2-j\xi k_0 - k^2/(3n^2)]}, \f where we use
the same notations as in \r{uniaxial_permittivity}, and $k^2 =
k_x^2+k_y^2+k_z^2$. The obtained permittivity dyadic can be also
rewritten as \e {\=\E(\o,\_k)\over \E_0} = \E_{\rm tr}(\o)\left(\=I -
  {\_k\_k\over k^2}\right)+ \E_{\rm lo}(\o,k){\_k\_k\over k^2}, \f
where
\begin{eqnarray}
\l{eps_t} \E_{\rm tr}(\o)    &=& \E_{\rm t} - {k_{\rm p}^2\over k_0^2-j\xi k_0},\\
\l{eps_l} \E_{\rm lo}(\o,k) &=& \E_{\rm t} - {k_{\rm p}^2\over k_0^2-j\xi k_0 - k^2/(3n^2)}.
\end{eqnarray}

It can be verified that for the mesh of thin plasmonic rods without
loading [for which $Z_w$ is given by  Eq. \r{Zw}], the relations
\r{eps_t}--\r{eps_l} transform to the ones presented in
Ref.~\onlinecite{Silveirinha_crosswires_PRB_2009} with the parameters $\E_{\rm
t} = 1$, $k_{\rm p} = \beta_{\rm p}$, and identifying the numerical
coefficient $l_0$ from the same reference with $l_0 = 3n^2$.

\section{Uniaxial wire medium loaded with patches and suppression of spatial dispersion \label{sect_uniax_supr}}
\label{supression} Recently\cite{Demetriadou_taming_JPCM_2008} it was
proposed to load the wire mesh with metal patches to increase the
effective capacitance of the wires per unit length and decrease the
related spatial dispersion effects. This proposal was supported by
numerical simulations. Here, we will apply our general analytical
model to the particular case of a uniaxial wire medium loaded with
metal patches. For this purpose we just need to determine what is
the effective capacitance $C$ introduced in Section~\ref{uniaxial}
in the presence of patches. The details of calculation of this
capacitance are described in Appendix A. Here we give the result: $C
= C_{\rm wire} + C_{\rm patch}$, where $C_{\rm
  wire}$ is the wire capacitance given by \r{capacitance} and \e
C_{\rm patch} = {2\pi\E_0w\over h\log\left(\sec{\pi d\over
      2a}\right)}, \l{cpatch} \f where $w$ is the width of the square
patches periodically attached to the wires and separated by the
distance $h$ along $z$, and $d = a - w$ is the gap between two
adjacent patches on a pair of neighboring wires. Thus, the
permittivity dyadic of the uniaxial wire medium loaded with patches
is given by \e {\=\E\over \E_0} = \E_{\rm
  t}\=I_{\rm t}+\left(1 - {k_{\rm p}^2\over k_0^2 - j\xi k_0 - k_z^2 /
    n^2}\right)\_z_0\_z_0, \l{unipatchperm} \f where we keep the same
notations as in Section~\ref{uniaxial}.  The transverse permittivity
$\E_{\rm t}$ is mostly determined by the patches when $w\gg r_0$ and
it can be found as the permittivity of a stack of capacitive grids
separated by $h$ one from another. With the help of the known theory
of such grids~\cite{Tretyakov_modelling_2003} it can be found that
\e \E_{\rm t} = 1 + {2w\over \pi h}\log\left(\csc{\pi d\over
    2a}\right).  \l{epst} \f The accuracy of \r{cpatch} and \r{epst}
is better for small gaps and for large values of $h/a$.

In the limit $d\rightarrow 0$ the effective capacitance behaves as
$C \approx {16\E_0wa^2\over \pi h d^2}$ and, therefore, can be
arbitrarily large if the gap between two adjacent patches is made
small enough. On the other hand, the transverse permittivity
$\E_{\rm t}$ grows under the same limit as $\E_{\rm t}\approx
{2w\over \pi
  h}\log\left({2a\over\pi d}\right)$. The square of the slow-wave
factor $n^2$ is proportional to the effective capacitance,
therefore, by increasing the width of the patches one can discard
the spatial dispersion term in the right-hand side of
\r{unipatchperm} while keeping $\E_{\rm t}$ at a reasonable level
(this is possible because $\E_{\rm t}$ grows more slowly when
$d\rightarrow 0$). An explicit expression for the slow-wave factor
under the mentioned limit reads \e n^2 = {LC\over\E_0\M_0} \approx 1
+ {16 w\over \pi h (k_{\rm p}d)^2}. \f In fact, we have numerically
checked that this simple expression works quite well for gaps
of width $d \le 0.2a$.

\begin{figure}[htb]
\centering \epsfig{file=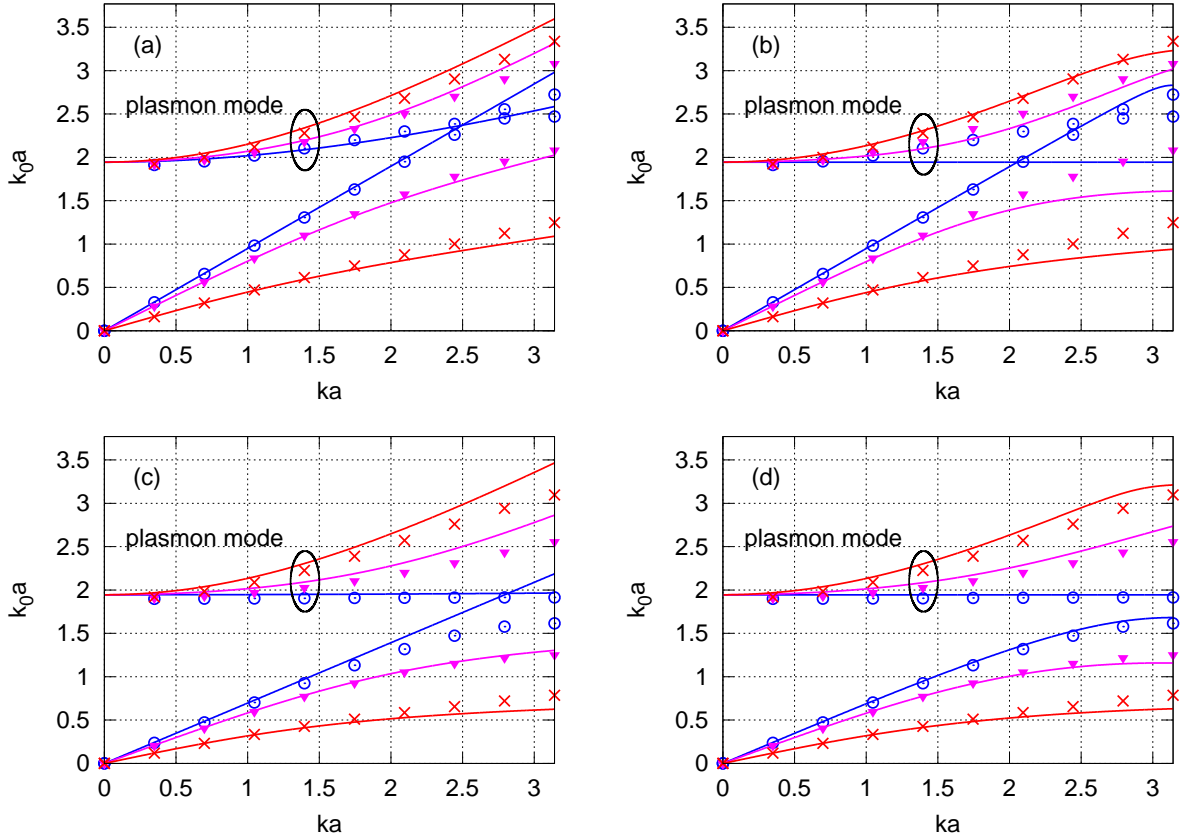,width=\textwidth}
\caption{\label{model_vs_cst} (Color online) Dispersion diagrams for
  a uniaxial wire medium loaded with patches obtained using
  two analytical models and numerical simulations for different
  propagation angles $\alpha$ with respect to the $z$-axis.
  Only the branches associated with the quasi-TEM and TM modes are shown.
  Panels (a)
  and (c): quasi-static model vs. numerical simulations: (a) $w =
  0.5a$, (c) $w = 0.9a$. Panels (b) and (d): transfer matrix model
  vs. numerical simulations: (b) $w = 0.5a$, (d) $w = 0.9a$. On all 4
  panels the solid lines represent the analytical results and the
  symbols correspond to the results of numerical simulations; the
  values of the propagation angles are coded in color: $\alpha=0$:
  blue lines and circles; $\alpha=30^\circ$: magenta lines and
  triangles; $\alpha=60^\circ$: red lines and crosses. The other
  parameters in all 4 cases: $r_0 = 0.05a$, $h = a$.}
\end{figure}

To illustrate the suppression of the spatial dispersion in the
considered wire media, we have calculated the dispersion diagrams
for several configurations using our quasi-static model, the
transfer matrix method described in Appendix B, and the eigenmode
solver of CST Microwave Studio. The structure was assumed lossless
in the simulations (all metallic components are perfectly conducting
so that $Z_w=0$). The transfer matrix formalism developed in
Appendix B is based on the assumption that in between two patch
grids the electric field is a superposition of TEM and TM modes.
\cite{Belov_dispersion_PRB_2003} The fields on the interfaces of
each patch grid are linked by a grid impedance and by an additional
boundary condition,\cite{ABCtilted} consistent with the formalism
described in Refs.~\onlinecite{Olli_Mushrooms_MTT_2009,
Alex_Mushrooms_MTT_2009}. The obtained results are presented in
Fig.~\ref{model_vs_cst}.

In Fig.~\ref{model_vs_cst}(a) and Fig.~\ref{model_vs_cst}(c) the
dispersion diagrams obtained from the quasi-static model and the
numerical simulations are shown for a set of the propagation angles
with respect to the axis of the structure: $\alpha = 0, 30^\circ,
60^\circ$ [for the other parameters of the structure refer to
Fig.~\ref{fig1}; in these plots the wave vector is ${\bf{k}} = k\left(
{\sin \alpha \,{\bf{x}}_0  + \cos \alpha \,{\bf{z}}_0 } \right)$].
The dispersion curves predicted by the quasi-static model are
depicted with solid lines while the results of the numerical
simulations are represented by symbols. In the example of
Fig.~\ref{model_vs_cst}(a) the patch width has been set equal to $w
= 0.5 a$, while in Fig.~\ref{model_vs_cst}(c) the patch width is $w
= 0.9a$. In both cases the theory and the simulations predict the
existence of two dispersion branches associated with extraordinary
waves, i.e., with the quasi-TEM and TM modes, as well as a dispersion
branch associated with the ordinary (TE) wave whose dispersion is
not depicted in Fig.~\ref{model_vs_cst} (there are also other higher
order modes at higher frequencies, but we are not interested in
them). We call the high-frequency branch ``the plasmon mode''
because for $\alpha = 0$ this branch corresponds to the longitudinal
plasmon-type wave propagating along the axis of the structure. On
the other hand, the low-frequency branch for $\alpha = 0$ belongs to
an ordinary transverse wave which is not affected by the wires (but
it is affected by the transverse permittivity $\E_{\rm t}$ of the
medium).

From Fig.~\ref{model_vs_cst}(a) one can see that for the
moderate-size patches the quasi-static model works surprisingly well
even when $ka$ approaches $\pi$.  The small difference in the
frequencies of the plasmon-type modes predicted by the theory and
the simulations at $ka = 0$ is due to the asymptotic nature of the
formula for the plasma wavenumber that we use (the discussion on
this is given in Section~\ref{uniaxial}). For larger patches
(Fig.~\ref{model_vs_cst}(c)) the quasi-static model does not predict
appearance of a band gap at $\alpha = 0$ and $ka = \pi$. This is
expected since in the model the capacitive loading on the wires is
assumed to be effectively uniform along the wires.

Fig.~\ref{model_vs_cst}(b) and Fig.~\ref{model_vs_cst}(d) display
the same dispersion diagrams but with the quasi-static model
replaced by the transfer matrix model described in Appendix B. One
can see that this model wrongly predicts a completely flat
dispersion for the plasmon mode propagating along the $z$ axis
($\alpha = 0$), independently of the patch size. This is in
disagreement with the numerical simulations, as is seen from
Fig.~\ref{model_vs_cst}(b). Indeed, the formalism developed in
Refs.~\onlinecite{Olli_Mushrooms_MTT_2009, Alex_Mushrooms_MTT_2009} is only
valid when the gap between the patches is small, because otherwise
other higher modes can be excited near the connections of the wires
to the patch grid, and in such conditions it is not possible to
consider that the microscopic field in the vicinity of the
connection points are a superposition of TM and TEM modes of the
unloaded wire medium, as assumed in
Refs.~\onlinecite{Olli_Mushrooms_MTT_2009, Alex_Mushrooms_MTT_2009}. Consistent
with this observation, it is seen in Fig.~\ref{model_vs_cst}(d),
that for larger patches and (or) larger angles of propagation the
disagreement is less pronounced. Another characteristic feature of
the transfer matrix model is that it is able to predict the
existence of the above-mentioned bandgap. This is because the
transfer matrix model takes into account the granularity of the
structure along the $z$ axis.

\begin{figure}[htb]
\centering
\epsfig{file=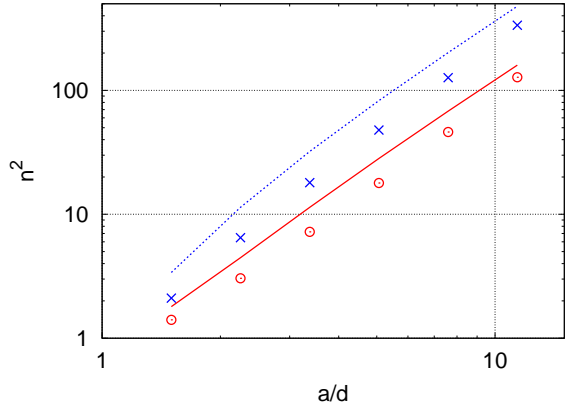,width=0.5\textwidth}
\caption{\label{slow_wave_factor} (Color online) The square of the
  slow-wave factor as a function of $a/d$ (logarithmic scale). The
  lines represent the result of the quasi-static model, the symbols
  correspond to the values of $n^2$ extracted from the numerical
  simulations. Blue dotted line and
  crosses: $r_0 = 0.05a$, $h = a/3$; red solid line and circles: $r_0 =
  0.05a$, $h = a$.}
\end{figure}

The suppression of the spatial dispersion effects is evident if we
compare Fig.~\ref{model_vs_cst}(a) with Fig.~\ref{model_vs_cst}(c).
Indeed, the latter case corresponds to a larger patch width ($w =
0.9a$), and consequently the slope of the dispersion curve
associated with the longitudinal mode (the plasmon mode at $\alpha =
0$) is very small. To justify this effect and also to check the
accuracy of the quasi-static model near the origin of the Brillouin
zone for a wide range of values of the gap, we have extracted the
values of the slow wave factor $n$ from the results of the numerical
simulations slightly above the point $ka = 0$ and compared them with
the value of $n$ given by the analytical model. The results of this
extraction are presented in Fig.~\ref{slow_wave_factor}. From this
figure we see that despite its simplicity, the quasi-static model
predicts very well the trend in the growth of $n^2$ when the the gap
between the patches decreases. The agreement tends to improve for
larger values of $h/a$.

\section{Conclusions}

In this paper we have developed a quasi-static analytical model of
wire media applicable to a wide class of structures, and in
particular we have considered uniaxial and isotropic wire crystals,
which may be loaded with metallic patches. Because the developed
model is defined in simple physical terms of the effective
inductance and capacitance per unit length of a wire it can be
readily extended to other wire structures of more complex
geometries.  The model accounts for the finite conductivity of the
wires so that it can be applied when the metallic wires become
plasmonic (consistent with the results reported in
Refs.~\onlinecite{Silveirinha_crosswires_PRB_2009,
Silveirinha_ENG_Plasmonic_2006}) or when the wires are uniformly
loaded with arbitrary complex impedances. In particular, we have
studied with details the electrodynamics of uniaxial wire media
loaded with patches, and demonstrated with full wave simulations
that the proposed quasi-static model describes accurately the
properties of the system in the long wavelength limit. Consistent
with the analysis of Refs.~\onlinecite{Demetriadou_taming_JPCM_2008,
Olli_Mushrooms_MTT_2009, Alex_Mushrooms_MTT_2009}, it was shown that
the presence of the patches may result in a dramatic reduction of
the nonlocal effects. For the case of unloaded wire media, we have
demonstrated that the quasi-static model yields the same expressions
for the dielectric permittivity tensors as those obtained by much
more sophisticated methods.\cite{Shvets_wires_PSPIE_2003,
Efros_WM_PRB_2002, Silveirinha_ENG_Plasmonic_2006,
Silveirinha_3dconnected_I3EMTT_2005, Constantin_WM,
Silveirinha_crosswires_PRB_2009} Thus, we have proven that the
strong spatial dispersion in wire media is a quasi-static effect.
Although this fact has already been noticed,~\cite{Maslovski_disser_2004} the presented research extends the
results obtained in Ref.~\onlinecite{Maslovski_disser_2004} and allows for
analytical and quantitative studies of the possibilities to control
the spatial dispersion in wire media.

\begin{acknowledgments}
This work is supported in part by Funda\c{c}\~ao para a Ci\^encia e
a Tecnologia under project PDTC/EEA-TEL/71819/2006.
\end{acknowledgments}

\appendix

\section{\label{ApA}}

As is seen from Fig.~\ref{fig1} (right) that depicts the path along
which we calculate the circulation of the electric field, the
capacitance in question can be calculated if we find the electric
field in the region close to the gap between two patches on the
neighboring wires. Indeed, the circulation integral \r{circint} in
the presence of patches has to be modified as follows
\begin{multline}
\oint\_E\.\_{dl} = \int\limits_{z_0}^{z_0+h}\!\!E_z(r_0, z)\,dz
-\!\!\int\limits_{z_0}^{z_0+h}\!\!E_z(a/2, z)\,dz
+\int\limits_{w/2}^{a/2}E_x(x,z_0+h)\,dx
-\int\limits_{w/2}^{a/2}E_x(x,z_0)\,dx, \l{circintA}
\end{multline}
where $z_0$ is at the location of an arbitrary plane of patches.
We choose the integration path so that it first goes along the
surfaces of the wire and the patch till the gap, then across the gap
till the symmetry line (Fig.~\ref{fig1}), then along that line till
the second gap and then across this gap back to the patch and the
wire.

One can see that the first integral in the right-hand side of~\r{circintA} still represents
the same quantity as in the unloaded uniaxial wire medium and is
related to the finite conductivity of the wire. The integrals along
the surfaces of the patches are not shown in \r{circintA} as in the
following we consider the patches to be ideally conducting. This may
be a good approximation because in practice the wire impedance
dominates. We can express the second integral of \r{circintA} as
\e
\int\limits_{z_0}^{z_0+h}\!\!E_z(a/2, z)\,dz =
\langle E_z\rangle h - \varphi_{z}(z_0 + h) + \varphi_{z}(z_0),
\l{phiz}
\f
where the two last terms account for the strong non-uniformity of the
$z$-component of the microscopic electric field in the vicinity of the
gaps and can be defined (in the unit cell $z_0 \le z \le z_0+h$) as
\e
\varphi_z(z) = \int\limits_z^{z_0+h/2}\!\!\!(E_z(a/2,z) - \langle E_z\rangle)\,dz.
\f
Notice that when there are no patches the microscopic field
changes smoothly along $z$; that is why in Section~\ref{uniaxial} we
could simply relate the second integral of~\r{circintA} with the macroscopic electric
field.  The two remaining integrals can be written in terms of \e
\varphi_x(z) = \int\limits_{w/2}^{a/2}E_x(x,z)\,dx.  \f

Substituting the above expressions for the integrals into \r{circintA} and comparing it with
\r{circulation} with $\Delta z = h$ we see that the additional potential at the plane $z=z_0$ has to be
\e \varphi(z_0) = \varphi_x(z_0) +
\varphi_z(z_0).
\l{avphi}
\f 

At this point it is worth reminding that the additional potential as we define it and use it in
Sections~\ref{uniaxial} and \ref{mesh} is essentially a macroscopic quantity: It changes slowly and smoothly along the wires.
Therefore, Eq.~\r{avphi} can be understood as the definition of the averaging procedure for the additional electric field
(represented by both non-uniform $E_x$ and $E_z$ components) that appears because of the periodical non-uniformity in the charge
distribution introduced by the patches.

It is clear that for wide patches this additional potential is mainly determined
by the fluctuating part of the microscopic field in the vicinity of the gap at $x = a/2$, $z = z_0$.
Thus, to simplify the problem we may first neglect the effect of the
charges sitting on the wires (nevertheless, we will later add a correction term taking the wires into account).
Second, because $\varphi(z_0)$ depends only on the non-uniform part of the field we may neglect the effect of all other planes of
charged patches except the plane $z = z_0$. We may do so because when $h \gg d = a - w$ the field produced
by the other planes of patches is practically uniform in the vicinity of the gap we are interested in.
Therefore in the following we consider only a single array of charged patches and discard all the patches
that are not co-planar with the patch at $z = z_0$.

In what follows, we will solve the enunciated electrostatic problem
and calculate the effective capacitance per unit length of a wire with patches.
In order to obtain a closed-form analytical solution we will make an additional simplification: we
replace all the patches centered at the same $y$-coordinate with a
single metal strip of the same width. Thus, we obtain a grid of
metallic strips (geometry of the problem becomes invariant along
$y$) separated by the same gap as the array of patches.

\begin{figure}[htb]
\centering
\epsfig{file=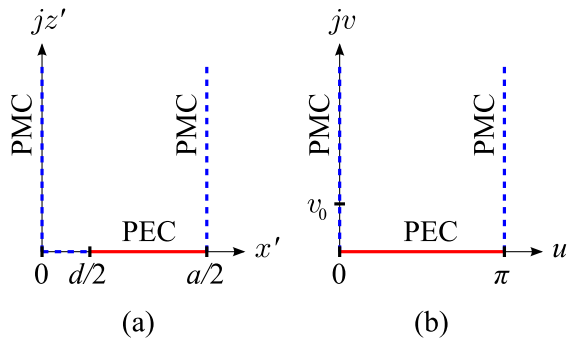,width=0.45\textwidth}
\caption{\label{mapping} (Color online) The original domain (a) and the domain
  obtained after the conformal mapping (b) defined by Eq.~\r{xi}. The
  solid red lines represent the perfect electric conductor boundary
  (PEC) (surface of the patch), the dashed blue lines represent the
  perfect magnetic conductor (PMC) boundaries that impose necessary
  symmetries. The point $v_0$ corresponds to the origin of the domain
  (a).}
\end{figure}

Taking into account the symmetry of the excitation and the
periodicity of the grid we arrive at the two-dimensional problem
shown in Fig.~\ref{mapping}(a). In this figure we define the local coordinate system $x'$,$z'$ as follows.
The middle point of the gap is at $x'=z'=0$. The patch is modeled as an infinitely thin
PEC (perfect electric conductor) strip, which starts at $x' =
d/2$ and continues to the point $x' = a/2$ which is at the middle
line of the patch.  The strip is charged. The PMC (perfect magnetic
conductor) boundaries shown in the figure enforce the symmetries
mentioned above.

The distribution of the electric potential in this system can be
found with the conformal mapping approach.\cite{Collin} One can
verify that the following analytical function of the complex
variable ${\cal Z} = x' + jz'$ \e \zeta({\cal Z}) = u(x',z') + j v(x',z') =
\cos^{-1}\left[{2\cos{2\pi {\cal
        Z}\over a} - \cos{\pi d\over a} + 1 \over \cos{\pi d\over a} +
    1}\right] \l{xi} \f maps the domain shown in Fig.~\ref{mapping}(a)
into the domain of Fig.~\ref{mapping}(b) in which the solution for the
electric field is trivial. One can see that the curves $v(x',z') =
\mbox{const}$ are the equipotential contours and the curves $u(x',z') =
\mbox{const}$ are the force lines of the electric field of the
problem. It can be checked that at the the patch the potential defined
in this way vanishes: $v(x',0) = 0$ for $d/2 \le x' \le a/2$.

Therefore, the voltage drop between a distant point on the $z'$ axis
(which is a point on the integration path shown in Fig.~\ref{fig1})
and the edge of the patch is \e v(0,z') =
\cosh^{-1}\!\left[{2\cosh{2\pi z'\over a}-\cos{\pi d\over a} + 1\over
    1+\cos{\pi d\over a}}\right].  \f
When $z'\gg d$ this voltage
asymptotically behaves as \e v(0,z') \sim 2\log\left(\sec{\pi d\over
    2a}\right) + {2\pi z'\over a}.  \l{volt} \f The linearly growing
term of \r{volt} corresponds to a uniform electric field far away
from the gap. Such a smooth field is already taken into account by
the first term of \r{phiz}. Therefore, the additional potential we
are looking for must be defined as \e \varphi(z_0) =
\lim_{z'\rightarrow\infty}\left(v(0,z') - {2\pi z'\over
    a}\right) = 2\log\left(\sec{\pi d\over 2a}\right).  \f

On the other hand, the total charge per unit length of the strip is
given by \e Q = 4\E_0[u(a/2,0)-u(d/2,0)]=4\pi\E_0, \f where the
coefficient $4$ accounts for the fact that the domain of
Fig.~\ref{mapping}(a) includes only $1/4$ of the total surface of
the strip.  Hence, the effective capacitance per unit length of a
strip is $C_{\rm strip} = Q/\varphi(z_0)$ and the effective patch
capacitance per unit length of a wire with patches is $C_{\rm patch}
= (w/h) C_{\rm strip}$, which is given by Eq.~\r{cpatch}. The total capacitance per
unit length of a wire with patches is approximated as a sum of the
wire capacitance $C_{\rm wire}$ from~\r{capacitance} and $C_{\rm
  patch}$.

\section{\label{ApB}}

The uniaxial wire medium periodically loaded with patches may be
regarded as a layered structure. Thus, it is possible to apply the
standard transfer matrix method to it, provided we are able to
characterize the fields in one cell. As demonstrated next, this can
be done by generalizing the formalism developed in
Refs.~\onlinecite{Olli_Mushrooms_MTT_2009, Alex_Mushrooms_MTT_2009}, which is
based on the assumption that in the regions in between two arbitrary
adjacent patch grids the microscopic fields can be written as a
superposition of TM and TEM modes of the unloaded wire medium.

Suppose that the wires are directed along $z$ and that the magnetic
field is along the $y$-direction (the TM polarization).  The
electric field components are $E_x$ and $E_z$ and the wave vector is
${\bf{k}} = k_x {\bf{x}}_0  + k_z {\bf{z}}_0$. The fields in the
region $0 < z < h$ ($z = 0$ at the patch array) in between two
arrays of metallic patches can be decomposed into four waves: \e
\eta _0 H_y = A_{\rm
  TM}^ + e^{ - \gamma _{\rm TM} z} + A_{\rm TM}^ - e^{ + \gamma _{\rm
    TM} z} + B_{\rm TEM}^ + e^{ - \gamma _{\rm TEM} z} + B_{\rm TEM}^
- e^{ + \gamma _{\rm TEM} z}, \l{mario_hy} \f
\begin{multline}
  E_x = \frac{j}{{\varepsilon_0 k_0 }}\frac{d}{{dz}}\left( {\eta _0
      H_y } \right) =
  -\frac{j}{{\varepsilon_0 k_0 }}\left[ \gamma _{\rm TM}\left( {A_{\rm TM}^ +  e^{ - \gamma _{\rm TM} z}  - A_{\rm TM}^ -  e^{ + \gamma _{\rm TM} z} } \right) + \right.\\
  \left.\gamma _{\rm TEM}\left( {B_{\rm TEM}^ + e^{ - \gamma _{\rm
            TEM} z} + B_{\rm TEM}^ - e^{ + \gamma _{\rm TEM} z} }
    \right) \right],
\end{multline}
\e E_z = \frac{{ - k_x }}{{\varepsilon _{zz}^{\rm TM} k_0 }}\left(
  {A_{\rm TM}^ + e^{ - \gamma _{\rm TM} z} + A_{\rm TM}^ - e^{ +
      \gamma _{\rm TM} z} } \right), \l{mario_ez} \f where
$\gamma_{\rm TM} = \sqrt{k_{\rm p}^2+k_x^2-k_0^2}$, $\gamma_{\rm TEM}
= jk_0$, and $\varepsilon_{zz}^{\rm TM} = \varepsilon_0k_x^2/(k_{\rm
  p}^2+k_x^2)$. It may be easily recognized that the above
  expressions correspond to a superposition of the standard TEM and
  TM modes supported by the unloaded wire medium.\cite{Belov_dispersion_PRB_2003}

The tangential components of the fields at the planes $z = h^-$ and
$z = h^+$ are linked by the transfer matrix of the patch array: \e
\left(
  {\begin{array}{*{20}c}
      {E_x }  \\
      {\eta _0 H_y }  \\
    \end{array}} \right)_{z = h^+  }  = \left( {\begin{array}{*{20}c}
      1 & 0  \\
      { - y_{\rm g} } & 1  \\
    \end{array}} \right)\cdot\left( {\begin{array}{*{20}c}
      {E_x }  \\
      {\eta _0 H_y }  \\
    \end{array}} \right)_{z = h^-  },
\l{mario_grid} \f where $y_{\rm g}  = 2j\varepsilon_0 \frac{{k_0a}}
{\pi } \log \left( {\csc \left( {\frac{{\pi d}}
        {{2a}}} \right)} \right)$ is the normalized effective admittance of the patch
array.\cite{Olli_Mushrooms_MTT_2009, Alex_Mushrooms_MTT_2009}
On the other hand, it was shown in Ref.~\onlinecite{ABCtilted} that since
the microscopic surface charge density must vanish at the connections between the
wires and the grid, the following additional boundary condition
(ABC) must be verified: \e k_0 \varepsilon_0 \frac{{d{\kern 1pt} E_z
}}{{dz}} + k_x \eta _0 \frac{{d{\kern 1pt} H_y }}{{dz}} = 0, \quad z
= 0^+, z=h^-, \f where $\eta_0$ is the free-space impedance. Using
this ABC in Eqs.~\r{mario_hy}--\r{mario_ez}, it is possible to
obtain the coefficients associated with the TEM wave
($B_{TEM}^{\pm}$) as a function of the coefficients associated with
the TM wave ($A_{TM}^{\pm}$).  Then, $A_{TM}^{\pm}$ can be expressed
in terms of the tangential electric and magnetic fields at the $z =
0^+$ plane. Proceeding in this manner, it is possible to obtain
after lengthy but straightforward calculations the following
transfer matrix relation for the layer of the wire medium in between
two patch grids: \e \left( {\begin{array}{*{20}c}
      {E_x }  \\
      {\eta _0 H_y }  \\

    \end{array} } \right)_{z = h^-  }  = {\rm \bf{M}} \. \left( {\begin{array}{*{20}c}
      {E_x }  \\
      {\eta _0 H_y }  \\

    \end{array} } \right)_{z = 0^+  },
\l{mario_matrix} \f where the matrix ${\rm \bf{M}}$ is \e {\rm
\bf{M}} = \left( {\begin{array}{*{20}c}
      {m_{11} } & {m_{12} }  \\
      {m_{21} } & {m_{11} }  \\

    \end{array} } \right),
\f with the elements given by the following formulas: \e m_{11}
 = \frac{{\left( {\varepsilon_0  - \varepsilon _{zz}^{\rm TM}
} \right)\gamma _{\rm TM} \sinh \left( {\gamma _{\rm TM} h}
\right)\cosh \left( {\gamma _{\rm TEM} h} \right) + \varepsilon
_{zz}^{\rm TM} \gamma _{\rm TEM} \cosh \left( {\gamma _{\rm TM} h}
\right)\sinh \left( {\gamma _{\rm TEM} h} \right)}} {{\left(
{\varepsilon_0  - \varepsilon _{zz}^{\rm TM} } \right)\gamma _{\rm
TM} \sinh \left( {\gamma _{\rm TM} h} \right) + \varepsilon
_{zz}^{\rm TM} \gamma _{\rm TEM} \sinh \left( {\gamma _{\rm TEM} h}
\right)}}, \f \e m_{12}  = \frac{1} {{k_0 }}\frac{{j\gamma _{\rm
TEM} \gamma _{\rm TM} \sinh \left( {\gamma _{\rm TM} h} \right)\sinh
\left( {\gamma _{\rm TEM} h} \right)}} {{\left( {\varepsilon_0  -
\varepsilon _{zz}^{\rm TM} } \right)\gamma _{\rm TM} \sinh \left(
{\gamma _{\rm TM} h} \right) + \varepsilon _{zz}^{\rm TM} \gamma
_{\rm TEM} \sinh \left( {\gamma _{\rm TEM} h} \right)}}, \f
\begin{multline}
  m_{21} = { - jk_0 }\left[\frac{2\left( {\varepsilon_0 - \varepsilon
          _{zz}^{\rm TM} } \right)\varepsilon _{zz}^{\rm TM} \left[ {
          - 1 + \cosh \left( {\gamma _{\rm TM} h} \right)\cosh \left(
            {\gamma _{\rm TEM} h} \right)} \right]}
    {{\left( {\varepsilon_0  - \varepsilon _{zz}^{\rm TM} } \right)\gamma _{\rm TM} \sinh \left( {\gamma _{\rm TM} h} \right) + \varepsilon _{zz}^{\rm TM} \gamma _{\rm TEM} \sinh \left( {\gamma _{\rm TEM} h} \right)}}+\right.\\
  \left.\frac{\sinh \left( {\gamma _{\rm TEM} h} \right)\sinh \left(
        {\gamma _{\rm TM} h} \right)\left[ {\left( {\varepsilon_0 -
              \varepsilon _{zz}^{\rm TM} } \right)^2 \frac{{\gamma
              _{\rm TM} }} {{\gamma _{\rm TEM} }} + \left(
            {\varepsilon _{zz}^{\rm TM} } \right)^2 \frac{{\gamma
              _{\rm TEM} }} {{\gamma _{\rm TM} }}} \right]} {{\left(
          {\varepsilon_0 - \varepsilon _{zz}^{\rm TM} } \right)\gamma
        _{\rm TM} \sinh \left( {\gamma _{\rm TM} h} \right) +
        \varepsilon _{zz}^{\rm TM} \gamma _{\rm TEM} \sinh \left(
          {\gamma _{\rm TEM} h} \right)}}\right].
\end{multline}
It may be verified that $\det(\bf{M})=\rm{1}$. From Eq.
\r{mario_matrix} and Eq. \r{mario_grid} it is clear that the global
transfer matrix of the system is:
\e
{\bf{M}}_{\rm g}  = \left( {\begin{array}{*{20}c}
   1 & 0  \\
   { - y_{\rm g} } & 1  \\
\end{array}} \right)\cdot{\bf{M}}.
\f As is well known, the dispersion characteristic of the
Bloch-Floquet modes supported by the periodic structure verifies $
\cos \left( {k_z h} \right) = {\rm{tr}}\left( {{\bf{M}}_{\rm g} }
\right)/2$, where ${\rm{tr}}\left( {...} \right)$ represents the
trace of the matrix. Thus, it follows that dispersion equation for
the Bloch-Floquet modes is: \e \cos \left(
  {k_z h} \right) = m_{11} - \frac{{y_{\rm g}}}{2}m_{12}, \f The transfer matrix dispersion diagrams
 calculated section \ref{sect_uniax_supr} were obtained using the
above equation.

%\bibliography{references}

\end{document}